\documentclass[sigconf,screen]{acmart}
\AtBeginDocument{%
  }

\usepackage{subcaption}
\usepackage{xspace}
\usepackage[most]{tcolorbox}
\usepackage{enumitem}
\usepackage{graphicx}
\usepackage{textcomp}
\usepackage[group-separator={,}, group-minimum-digits={3}]{siunitx}
\usepackage{multirow}
\usepackage{pifont}
\usepackage{diagbox}
\usepackage{verbatim}
\usepackage{tabularray}
\usepackage{natbib}
\setcitestyle{numbers, super}
\usepackage{colortbl}
\usepackage{listings}
\PassOptionsToPackage{usenames, dvipsnames}{xcolor}
\usepackage[dvipsnames]{xcolor}
\definecolor{gree}{HTML}{1e8449}
\definecolor{orang}{HTML}{b35809}
\usepackage{balance}
\usepackage{microtype}
\usepackage{url}

\definecolor{orang}{HTML}{b35809}

\newcommand{\RomanNum}[1]{\MakeUppercase{\romannumeral #1}}
\setcopyright{acmlicensed}
\copyrightyear{2018}
\acmYear{2018}
\acmDOI{XXXXXXX.XXXXXXX}
\acmConference[Conference acronym 'XX]{Make sure to enter the correct
  conference title from your rights confirmation email}{June 03--05,
  2018}{Woodstock, NY}
\acmISBN{978-1-4503-XXXX-X/2018/06}

\renewcommand\footnotetextcopyrightpermission[1]{} 
\begin{document}



\title{AutoEmpirical: LLM-Based Automated Research for Empirical Software Fault Analysis}

\author{Jiongchi Yu}
\email{jcyu.2022@phdcs.smu.edu.sg}
\authornote{Both authors contributed equally to this research.}
\orcid{0000-0002-2888-4499}
\affiliation{%
  \institution{Singapore Management University}
  \country{Singapore}
}

\author{Weipeng Jiang}
\email{lenijwp@stu.xjtu.edu.cn}
\orcid{0000-0002-0382-6401}
\authornotemark[1]
\affiliation{%
  \institution{Xi'an Jiaotong University}
  \country{China}
}

\author{Xiaoyu Zhang}
\email{joshiningrain@gmail.com}
\orcid{0000-0001-7010-6749}
\affiliation{%
  \institution{Nanyang Technological University}
  \country{Singapore}
}

\author{Qiang Hu}
\email{qianghu@tju.edu.cn}
\orcid{0000-0002-8251-1669}
\affiliation{%
  \institution{Tianjin University}
  \country{China}
}

\author{Xiaofei Xie}
\email{xfxie@smu.edu.sg}
\orcid{0000-0002-1288-6502}
\affiliation{%
  \institution{Singapore Management University}
  \country{Singapore}
}

\author{Chao Shen}
\email{chaoshen@xjtu.edu.cn}
\orcid{0000-0002-6959-0569}
\affiliation{%
  \institution{Xi'an Jiaotong University}
  \country{China}
}



\begin{abstract}
Understanding software faults is essential for empirical research in software development and maintenance. However, traditional fault analysis, while valuable, typically involves multiple expert-driven steps such as collecting potential faults, filtering, and manual investigation. These processes are both labor-intensive and time-consuming, creating bottlenecks that hinder large-scale fault studies in complex yet critical software systems and slow the pace of iterative empirical research.

In this paper, we decompose the process of empirical software fault study into three key phases: (1) research objective definition, (2) data preparation, and (3) fault analysis, and we conduct an initial exploration study of applying Large Language Models (LLMs) for fault analysis of open-source software. Specifically, we perform the evaluation on 3,829 software faults drawn from a high-quality empirical study. Our results show that LLMs can substantially improve efficiency in fault analysis, with an average processing time of about two hours, compared to the weeks of manual effort typically required. We conclude by outlining a detailed research plan that highlights both the potential of LLMs for advancing empirical fault studies and the open challenges that required be addressed to achieve fully automated, end-to-end software fault analysis.


\end{abstract}

\maketitle

\section{Introduction}

Empirical studies of software faults have long been invaluable to both academia and industry. By mining software repositories and analyzing real-world bugs, such studies uncover common fault patterns and inform improved development practices. However, conducting software fault analyses is notoriously labor-intensive and time-consuming. Researchers typically must manually collect bug data and label each fault according to taxonomies through multiple rounds of discussion, which often requires weeks of effort and scarce domain-specific expertise~\cite{dogga2023autoparts,herbold2022problems}. This heavy reliance on human labor creates a significant bottleneck: large and complex software systems often generate a massive volume of issues, yet small research teams cannot feasibly complete the required labeling work within reasonable time or resource limits. The prolonged study cycles ultimately reduce the timeliness and practical impact of empirical software research.

Recent advances in large language models (LLMs) offer a promising avenue to automate and accelerate empirical studies of software faults. Modern LLMs such as GPT-4 and Claude demonstrate remarkable capabilities~\cite{bubeck2023sparks,nori2023capabilities} in understanding both natural language and code, achieving strong performance in tasks including code generation, documentation, and bug fixing. These models can also act as intelligent agents capable of reading, reasoning, and writing~\cite{achiam2023gpt,lu2024ai}. Recent studies further show that LLMs can assume many tasks traditionally performed by human researchers in domains such as systematic literature reviews~\cite{khraisha2023can}, and research evaluation~\cite{di2025evaluation,boiko2023autonomous}. Building on this momentum, the concept of autoresearch~\cite{liu2025vision,hu2025survey,chen2025ai4research} has emerged, where researchers adapt automation frameworks powered by LLMs to their own domain-specific tasks.

This motivates us to consider how to harness LLMs to advance software engineering research. We argue that empirical studies of software faults are particularly well-suited for LLM-enabled auto-research. Such studies typically follow a well-defined workflow, including objective formulation, data preparation, taxonomy-based labeling, and comparative analysis, with only modest, domain-bounded adaptations across projects. Compared with highly technical research that hinges on original algorithmic ideas, fault studies place less emphasis on radical novelty and more on careful, consistent analysis and comparison, which aligns with the strengths of LLMs as meticulous readers and evaluators~\cite{zheng2023judging}. 


Therefore, to investigate whether LLMs can execute an entire empirical study on software faults with minimal human intervention, and to assess how well current foundation models perform in these roles, we propose the first pipeline for empirical fault studies with LLMs in the loop. Specifically, we decompose the research process into three well-established phases commonly adopted in prior works~\cite{quan2022towards,chen2023toward,chen2021empirical}: research definition, data preparation, and interactive fault analysis. We then assign each phase to collaborating LLMs, with each model responsible for executing the core research tasks within its phase, and systematically evaluate their outputs with expert-established ground truth.

To assess the feasibility of our LLM-driven pipeline, we replicate an existing empirical fault study using five state-of-the-art foundation models and find that the pipeline completes in about two hours, yielding a 20× speed-up over manual analysis. While LLMs achieve roughly 70\% precision in identifying fault-related issues, their classification accuracy into predefined taxonomies remains around 50\%, highlighting the need for stronger reasoning capabilities. Our study provides the first benchmark for automated software fault research.



\section{Background and Related Works}

\begin{figure*}[ht]
    \centering
    \includegraphics[width=0.9\linewidth]{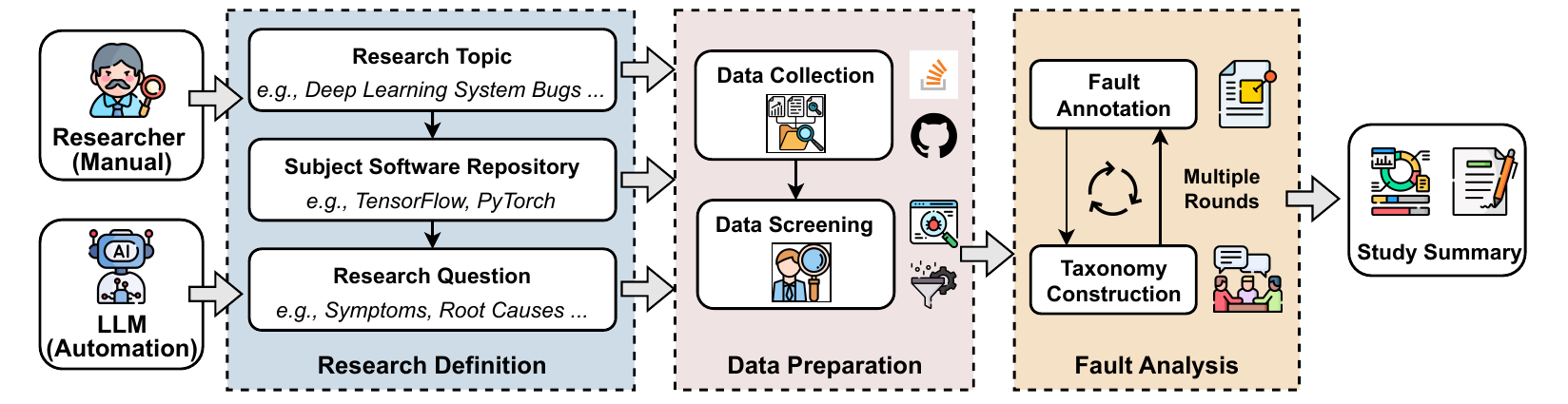}
    \Description{A typical workflow of empirical studies on software faults.}
    \caption{A typical workflow of empirical studies on software faults. }
    \label{fig:overview}
\end{figure*}


\textbf{Empirical Fault Study.} Empirical studies of software faults have long been a cornerstone of software engineering research. By manually analyzing real faults, researchers have uncovered recurring bug patterns and failure modes that inform both developers and tool builders. For example, Xiong et al.~\cite{xiong2023empirical} investigated hundreds of Android app failures, characterizing their root causes and required test oracles. Others have probed faults in emerging domains like machine learning software, for instance, Wang et al.~\cite{wang2022empirical} analyzed 400 TensorFlow and PyTorch faults and found prevalent numerical instability errors. We observe that established empirical software fault studies~\cite{quan2022towards,chen2021evaluating,lu2025empirical,yu2024bugs} typically follow a highly structured workflow comprising recurring phases: after defining the research questions and scope, the authors will collecting and filtering candidate issues, interatively annotating faults to form a taxonomy of fault characteristsic (such as symptoms), and finally synthesizing results into actionable insights. The process is dominated by mechanical, manual labeling and discussion tasks, often requiring weeks of expert effort. As a result, the scale and timeliness of empirical fault studies remain severely constrained.


\textbf{LLM in Automated Research.} 
The rise of LLMs has renewed interest in automating software engineering workflows. Modern code-aware LLMs bridge natural language and code, enabling summarization, explanation, and synthesis. Early tools like Copilot~\cite{chen2021evaluating} show their ability to generate correct code, foreshadowing applications in fault analysis and repair. More recent work demonstrates deeper reasoning: on SWE-bench, GPT-4 with tool support can interpret bug reports, navigate code, and propose patches~\cite{jimenez2023swe}. LLM-based pipelines also automate tasks such as untangling vulnerability fixes~\cite{wang2024reposvul} or detecting tangled commits with near-specialized accuracy~\cite{opu2025llm}. 
Beyond software engineering, LLMs are applied to large-scale research automation, from literature surveys~\cite{wang2024autosurvey,liang2025surveyx} to evaluating reviewer reports and paper quality~\cite{di2025evaluation,boiko2023autonomous}.
Building on this momentum, the concept of autoresearch is emerging: researchers adapt automation frameworks powered by LLMs to orchestrate domain-specific research tasks end to end ~\cite{liu2025vision,hu2025survey,chen2025ai4research}. In such systems, LLMs take on roles of literature retrieval, topic clustering, hypothesis generation, and draft synthesis, while humans act as overseers at the meta-level, guiding direction, validating results, and resolving ambiguity. Despite these advances, autoresearch remains unexplored in empirical software engineering, particularly for fault studies. We position our work as one of the first to apply the autoresearch paradigm to empirical fault analysis.




\section{Automated Software Fault Analysis}

As illustrated in~\autoref{fig:overview}, our approach decomposes the task of software fault analysis into three sequential stages. For each stage, we precisely define the required inputs, constraints, and the roles of LLM agents, enabling clear comparison between human and LLM-driven processes.

\subsection{Stage \RomanNum{1}: Research Definition}

End-to-end automated fault analysis begins with defining research objectives, typically informed by human expert requirements. LLMs can then assist in selecting appropriate repositories and determining analysis scope based on resource constraints. 
To preliminarily assess the capability of existing LLMs under the given research topics for empirical studies on software faults,
we query state-of-the-art models with targeted themes (e.g., fault studies of JavaScript engines) and task them with automatically selecting representative projects for fault analysis. We then compare the project selections made by the models against those chosen in established empirical studies. In addition, we prompt LLMs to autonomously formulate relevant research questions given project information and broader study contexts. This enables us to directly evaluate how effectively current foundation models can conceptualize and structure empirical studies on software faults.

\subsection{Stage \RomanNum{2}: Fault-Related Issue Selection}

After identifying the repositories for the study, Stage \RomanNum{2} focuses on selecting fault-related issues or commits from the repositories for further in-depth analysis. The key challenge is distinguishing these from typical feature-based commit records or non-fault issues.

\textbf{Input.} We begin with the complete collection of candidate issues (or commit records) from the target projects, including metadata such as title, state, timestamps, body text, and all associated comments. These raw records serve as the input to the LLM-based filtering pipeline.

\textbf{LLM Tasks.} For each candidate record, the LLM agent makes a binary decision: fault-related (eligible for further fault analysis) or non-fault. This judgment relies solely on the textual and metadata content of the issue/commit, under the assumption that no external code context is available at this stage.


\textbf{Filtering Criteria.} Instead of requiring models to design their own heuristics, the criteria are explicitly given by researchers to ensure comparability with prior work, while the judgment of whether individual issues satisfy these criteria is left entirely to the LLMs. As an example, the criteria used in our evaluation about DL library faults (\autoref{sec:experiment}) are encoded into the prompts as follows:


\begin{enumerate}
\item \textbf{DL-Relevance Requirement}: Issues must contain at least one keyword from the 147-word deep learning vocabulary, including TensorFlow.js-specific terms (e.g., "dispose", "WebGL") and general DL terminology.

\item \textbf{Actual Fault Reporting}: Issues must describe observable problems, errors, or system failures rather than feature requests, general questions, or theoretical discussions.

\item \textbf{Technical Clarity}: Issues must provide sufficient technical detail and clear problem descriptions that enable fault analysis and understanding.

\item \textbf{Exclusion of Non-Technical Content}: Issues with exclusion labels (e.g., "stat:awaiting response"), those without answers, or those related to deprecated versions (before 2020-01-01) are filtered out.
\end{enumerate}

\subsection{Stage \RomanNum{3}: Fault Taxonomy}

After selecting fault-related issues in Stage \RomanNum{2}, Stage \RomanNum{3} aims to classify each issue’s symptom and root cause using a predefined taxonomy. The challenge here is to test whether LLMs can perform fine-grained, semantically consistent classification that aligns with expert annotations.
Requesting an LLM to generate a full taxonomy from scratch poses significant alignment challenges: even if the model internally captures hierarchical structure, its labels or naming choices may diverge from human-designed ontologies. Prior work on automated taxonomy alignment underscores this divergence problem and explores techniques for reconciling mismatches between model-generated and expert taxonomies~\cite{cui2024automated}. To avoid this alignment barrier, we adopt a taxonomy-anchored classification strategy: we provide LLMs with complete taxonomic frameworks and definitions a priori, and then evaluate their ability to map textual inputs into this fixed taxonomy.



\textbf{Taxonomy-based Prompt Construction.} We systematically convert the taxonomies from the original literature into clear, structured prompts that provide LLMs with comprehensive classification frameworks. These prompts include:

\begin{itemize}
\item \textbf{Complete Hierarchical Structure}: Both symptom and root cause taxonomies with primary categories (Level 1), subcategories (Level 2), and specific types (Level 3)
\item \textbf{Detailed Definitions}: Precise descriptions for each category and subcategory, including characteristic indicators and distinguishing features
\item \textbf{Structured Output Format}: Standardized JSON format for consistent result collection and analysis
\end{itemize}

The symptom taxonomy encompasses five primary categories (Crash, Poor Performance, Build \& Initialization Failure, Incorrect Functionality, Document Error), with 15 subcategories and 15 leaf-level specific types. The root cause taxonomy includes five primary categories (Incorrect Programming, Configuration \& Dependency Error, Data/Model Error, Execution Environment Error, Unknown), with 17 subcategories from our selected literature~\cite{quan2022towards}.

\textbf{Fine-grained Evaluation Approach.} To comprehensively assess the fine-grained classification capabilities of LLMs, we use all leaf node labels as ground truth for evaluation. This approach ensures examination at the most granular level of classification accuracy, testing whether LLMs can discern subtle distinctions within the same hierarchical category.

For symptom classification, our leaf-level evaluation encompasses the finest taxonomic distinctions, such as differentiating between "DL Operator Exception" and "Function Inaccessible" within the Reference Error subcategory, or distinguishing "Memory Leak from "Out of Memory" within memory-related performance issues. For root cause classification, the evaluation focuses on subcategory-level precision, examining LLMs' ability to discriminate between closely related causes such as "Unimplemented Operator" versus "Inconsistent Modules in TF.js" within the broader Incorrect Programming category. This fine-grained evaluation framework rigorously tests the depth of LLMs' taxonomic understanding and their capacity for classification that mirror expert-level analysis.

\section{Preliminary Experiment}

\subsection{Experiment Setup}
\label{sec:experiment}

\begin{figure*}[t]
  \centering
  \begin{subfigure}{0.32\linewidth}
    \centering
    \includegraphics[width=\linewidth]{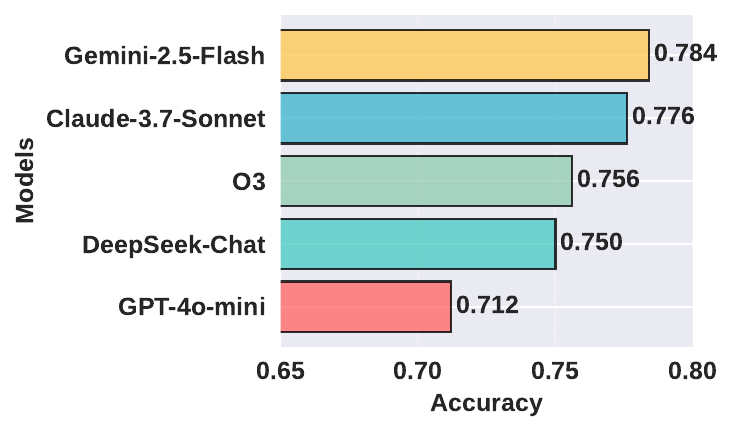}
    \caption{Bug-related Issue Filtering}
    \label{fig:rq2_performance}
  \end{subfigure}
  \hfill
  \begin{subfigure}{0.32\linewidth}
    \centering
    \includegraphics[width=\linewidth]{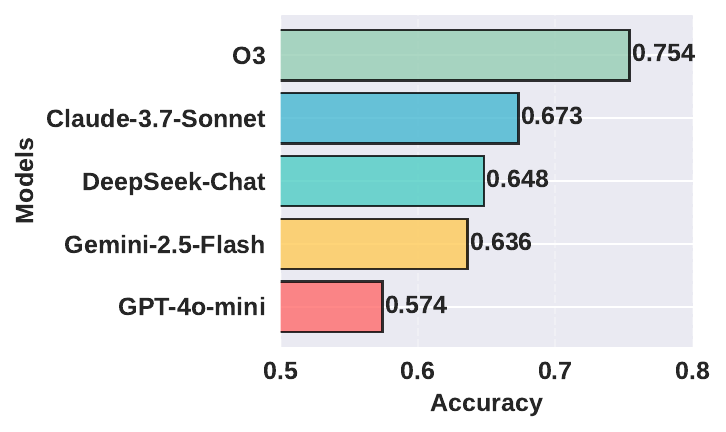}
    \caption{Bug Symptoms Classification}
    \label{fig:rq3_symptom_performance}
  \end{subfigure}
  \hfill
  \begin{subfigure}{0.32\linewidth}
    \centering
    \includegraphics[width=\linewidth]{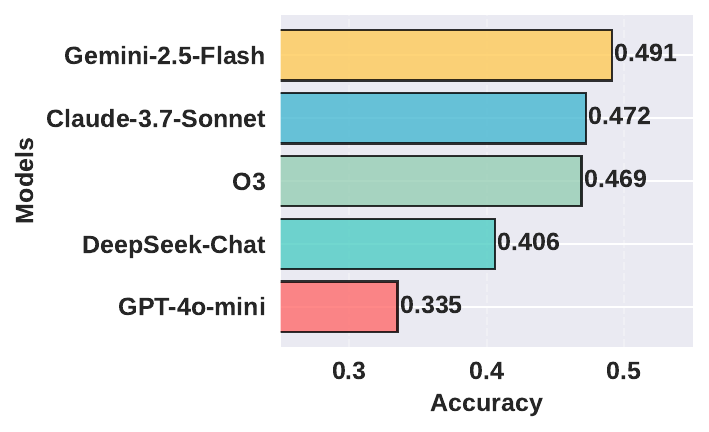}
    \caption{Bug Root Causes Classification}
    \label{fig:rq3_rootcause_performance}
  \end{subfigure}

  \caption{Overall classification performance on different stages.}
  \Description{Three side-by-side subfigures: filtering, symptoms, and root causes classification performance.}
  \label{fig:rq_performance}
\end{figure*}

\textbf{Dataset Construction for Stage \RomanNum{2}.}
Due to the computational cost of processing the full set of 3,859 candidate issues, we follow the methodology of the reference study~\cite{quan2022towards} and build a balanced evaluation subset by sampling 250 fault (positive) and 250 non-fault (negative) cases—500 in total. This stratified sampling ensures both evaluation tractability and class balance, enabling robust comparison of LLM performance in distinguishing fault-related issues. The primary evaluation metric for this task is how accurately LLMs replicate the expert filtering decisions from the original study.

\textbf{Ground Truth in Stage \RomanNum{3}.}
For the classification of symptoms and root causes, we adopt the full set of 684 annotated issues used in the original empirical study. Each issue in this dataset is labeled with expert-annotated symptom and root cause categories, offering a comprehensive and trustworthy ground truth. Because this set covers the full taxonomy of JavaScript DL system faults, it allows us to evaluate classification performance across all categories.

\textbf{Foundation Models.}
We evaluate five state-of-the-art LLMs, namely GPT-4o, GPT-o3, Claude 3.7, Gemini-2.5, and DeepSeek-V3, across our experiment in Stage \RomanNum{2} and Stage \RomanNum{3}. In line with prior work, we maintain consistency in data sampling and labeling rather than replicating every project in the literature~\cite{quan2022towards}.

\textbf{Metrics.}
In this work, we primarily report accuracy as the key metric, especially for multi-class classification tasks like fault taxonomy labeling, because accuracy provides a straightforward, aggregate measure of model correctness. Reporting accuracy allows clear comparison across models and tasks, while avoiding complications that arise in interpreting precision/recall trade-offs in imbalanced settings. 

\subsection{Performance of Stage \RomanNum{1}}

We task the LLMs with automatically selecting representative open-source projects relevant to the targeted research theme (i.e., JavaScript-based DL system faults). The models then formulate appropriate research questions based on these selections. We find that the LLMs consistently identify key projects such as \textit{TensorFlow.js} and also list several additional projects, including \textit{Teachable Machine} and \textit{Magenta.js}.
However, they tend to overlook third-party DL libraries and the broader set of all 58 JavaScript-based DL applications available for evaluation. This suggests that the LLMs exhibit a certain myopia, focusing primarily on repository-based cues and thus limiting the comprehensiveness of their project selection for related studies.

\subsection{Performance of Stage \RomanNum{2}}


As shown in~\autoref{fig:rq2_performance}, we find that all LLMs achieve promising performance in identifying bug-related issues, with Gemini-2.5-Flash demonstrating the best results. Most bug-related issues are successfully identified; however, the LLMs tend to include a greater number of non-security-related issues (such as testing issues), which leads to decreased accuracy.

\subsection{Performance of Stage \RomanNum{3}}






For the last stage~\footnote{Automated paper writing is an important research topic in its own right, encompassing literature retrieval, manuscript drafting, and figure generation, and has been discussed extensively in prior work. Therefore, it lies outside the scope of this study; ideally, the fault-analysis outputs in Stage \RomanNum{3} can be integrated with paper-generation tools to produce the final manuscript.}, as shown in~\autoref{fig:rq3_symptom_performance} and~\autoref{fig:rq3_rootcause_performance}, we are surprised to find that LLMs generally struggle to identify the root causes of bugs, even when provided with a predefined taxonomy. While Gemini demonstrates the best performance, it is able to correctly identify the root causes for only 49\% of the bugs. This limitation may be attributed to the lack of detailed code and contextual information within the issue reports, which would otherwise serve as essential guidance for LLMs in root cause analysis. This observation also highlights the need to design subsequent LLM agents with improved access to such contextual information. 

Additionally, we find that OpenAI GPT-4o exhibits outstanding performance in identifying bug symptoms, substantially outperforming the other models. In contrast, OpenAI GPT-4o-mini delivers the weakest performance among all evaluated LLMs, which aligns with results observed in mainstream benchmarks~\cite{perez2025ai4math,de2024show}.

\section{Future Plans}

Our preliminary investigation reveals both the potential and limitations of LLM-driven automation for empirical fault analysis. To advance this research toward robust, real-world applicability, we present a research agenda identifying key directions and challenges for future work.

\textbf{Comprehensive Dataset Collection.}
A fundamental requirement for rigorous evaluation is the availability of diverse, high-quality benchmarks. Future work focuses on systematically collecting and replicating datasets from empirical studies published in top-tier software engineering venues over the past five years. These studies typically release replication packages with analyzed faults, providing high-quality ground truth. The resulting benchmark will enable robust evaluation across diverse fault types and serve as a standardized testbed for future research.

\textbf{Evaluation of LLM-Based Autonomous Taxonomy Synthesis.} Our current evaluation pipeline relies on a pre-defined taxonomy, yet a fully automated agent should ideally generate and refine its own analytical frameworks. A key next step is to evaluate LLMs' capability to autonomously construct, apply, and refine fault taxonomies through expanded datasets and human studies. We will also explore multi-agent collaboration paradigms where specialized agents debate to reconcile differing classifications.
    
\textbf{Towards Fully Automated End-to-End Agent System.} The ultimate goal is a fully integrated system executing entire empirical workflows with minimal intervention. Achieving this vision presents several interconnected research challenges and opportunities: (1) \textit{Context-Aware Reasoning}: Current models struggle with root cause analysis due to insufficient code context. Future systems must dynamically retrieve and reason over source code, commit histories, and documentation. (2) \textit{Tool Integration and Agency}: Agents must proficiently use software engineering tools—static analyzers, debuggers, version control systems—to gather evidence and validate hypotheses. (3) \textit{Evolving Autonomy}: Systems must recognize limitations, identify ambiguous cases, and proactively seek clarification through iterative investigation.

\section{Conclusion}

In this paper, we propose the first fully automated empirical studies on the software fault pipeline. Our preliminary evaluation shows significant efficiency gains in fault data identification, though fine-grained taxonomy classification achieves only ~50\% accuracy. These results reveal both the potential and limitations of current LLMs for automated fault analysis. Our work establishes a foundation for scalable automated studies and identifies clear directions for improving LLM-driven research automation.



\bibliographystyle{ACM-Reference-Format}
\bibliography{icse26}

\end{document}